\documentclass[graybox]{svmult}

\usepackage{mathptmx}       
\usepackage{helvet}         
\usepackage{courier}        
\usepackage{type1cm}        
\usepackage{makeidx}         
\usepackage{graphicx}        
\usepackage{multicol}        
\usepackage[bottom]{footmisc}

\makeindex             

\begin{document}

\title*{Nonlinear Localization in Metamaterials}
\author{N. Lazarides and G. P. Tsironis}
\institute{N. Lazarides \at
Department of Physics, University of Crete, P. O. Box 2208, 71003 Heraklion, Greece
$\&$
Institute of Electronic Structure and Laser,
Foundation for Research and Technology-Hellas, P.O. Box 1527, 71110 Heraklion, Greece, 
\email{nl@physics.uoc.gr}
\and G. P. Tsironis \at 
Department of Physics, University of Crete, P. O. Box 2208, 71003 Heraklion, Greece
$\&$
Institute of Electronic Structure and Laser,
Foundation for Research and Technology-Hellas, P.O. Box 1527, 71110 Heraklion, Greece, 
\email{gts@physics.uoc.gr}}
%
%
\maketitle
\abstract*{
Metamaterials, i.e., artificially structured ("synthetic") media comprising weakly 
coupled discrete elements, exhibit extraordinary properties
and they hold a great promise for novel applications including super-resolution imaging, 
cloaking, hyperlensing, and optical transformation. Nonlinearity adds a new degree of 
freedom for metamaterial design that allows for tunability and multistability,
properties that may offer altogether new functionalities and electromagnetic
characteristics. The combination of discreteness and nonlinearity may lead to 
intrinsic localization of the type of discrete breather in metallic, SQUID-based,
and ${\cal PT}-$symmetric metamaterials.
We review recent results demonstrating the generic appearance of breather excitations  
in these systems resulting from power-balance between intrinsic losses and input
power, either by proper initialization or by purely dynamical procedures.
Breather properties peculiar to each particular system are identified and discussed.
Recent progress in the fabrication of low-loss, active and superconducting metamaterials,
makes the experimental observation of breathers in principle possible with the
proposed dynamical procedures.
}

\section{Introduction}
\label{sec:0}
Advances in theory and nanofabrication techniques have opened 
new unprecedented opportunities for researchers to create artificially structured
media with extraordinary properties that rely on particular geometric
arrangements. A well-known paradigm is that of {\em metamaterials} that provide 
access to all quadrants of the real permittivity-permeability plane, exhibiting 
negative refraction index, optical magnetism, and other fascinating properties
\cite{Shalaev2007,Soukoulis2007,Soukoulis2011,Zheludev2012}.
Their unique properties are particularly well suited for novel devices like 
hyperlenses \cite{Pendry2000} and optical cloaks of invisibility \cite{Schurig2006},
while they may form a material base for other functional devices with tuning and 
switching capabilities \cite{Zheludev2010,Zheludev2011}.
The key element for the construction of metamaterials has customarily been the 
split-ring resonator (SRR), a subwavelength resonant "particle" which is 
effectively a kind of an artificial "magnetic atom" \cite{Caputo2012}.
A periodic arrangement of SRRs in space forms a {\em magnetic metamaterial}
that exhibits high frequency magnetism and negative permeability \cite{Linden2006}.
In several applications, real-time tunability of the effective parameters of a 
metamaterial is a desired property, that can be achieved by nonlinearity
\cite{Powell2007,Shadrivov2008,Wang2008}.

Metamaterials comprising metallic elements suffer from high losses at frequencies 
close to those in their operating region, that place a strict limit on their 
performance and hamper their use in devices. The quest for loss compensation is 
currently following two different pathways: a "passive" one, where the metallic 
elements are replaced by superconducting ones \cite{Anlage2011}, and an "active"
one, where appropriate constituents are added to metallic metamaterials that 
provide gain through external energy sources. The latter has been recently recognized
as a very promising technique for compensating losses \cite{Boardman2010}.
{\em Superconducting metamaterials} exhibit both significantly reduced losses
and intrinsic nonlinearities due to the extreme sensitivity of the 
superconducting state to externally applied fields 
\cite{Ricci2005,Ricci2007,Gu2010,Fedotov2010,Chen2010}.   
The fabrication of superconducting SRRs with narrow slits filled with a dielectric 
oxide brings the Josephson effect into play \cite{Josephson1962}. For a thin
enough dielectric barrier a Josephson junction (JJ) is formed, and the currents
in the ring are then determined by the Josephson relations \cite{Josephson1962}.
The Josephson element thus turns the superconducting ring into an rf SQUID
(Superconducting QUantum Interference Device) \cite{Barone1982,Likharev1986},
a long known device in the Josephson community. 
The replacement of metallic and/or superconducting SRRs with rf SQUIDs, suggested
a few years ago \cite{Lazarides2007,Lazarides2008}, results in (SQUID-based) 
metamaterials with both reduced losses and yet another source of nonlinearity
due to the Josephson element. Thin-film metasurfaces using JJs as basic elements
has been recently demonstrated \cite{Jung2013}.

Nonlinear metallic metamaterials can be constructed by appropriate combinations
of highly conducting SRRs with nonlinear electronic components; several types of 
diodes have been successfully employed for this purpose 
\cite{Powell2007,Shadrivov2008,Wang2008}. In order to construct nonlinear and 
{\em active metamaterials}, however, gain-providing electronic components such
as tunnel (Esaki) diodes \cite{Esaki1958}, have to be employed. 
The latter feature a negative resistance part in their current-voltage
characteristics, and therefore can provide both gain and nonlinearity in an 
otherwise conventional metamaterial. Tunnel diodes may also be employed for the
construction of ${\cal PT}-$symmetric metamaterials, that rely on balanced gain 
and loss, in a way similar to that used in electronic circuits \cite{Schindler2011}.
${\cal PT}-$symmetric systems do not obey separately the parity ($\cal P$) and 
time ($\cal T$) symmetries, but instead they do exhibit a combined ${\cal PT}$ 
symmetry. The notions of ${\cal PT}-$symmetric systems originate for non-Hermitian 
quantum mechanics \cite{Hook2012}, but they have been recently extended to dynamical 
lattices, particularly in optics \cite{ElGanainy2007,Makris2008}.
Following these ideas, a ${\cal PT}$ metamaterial with elements having alternatingly
gain and equal amount of loss has been suggested \cite{Lazarides2013,Tsironis2013}.

Conventional (metallic), SQUID-based, and ${\cal PT}-$metamaterials 
share a number of common features. They can all be constructed by discrete 
elements which are weakly coupled through magnetic and/or electric forces
\cite{Sydoruk2006,Hesmer2007,Sersic2009,Rosanov2011}, while in most cases
the inter-element coupling may be limited to nearest-neighbors. 
SQUIDs are coupled magnetically; also, for particular mutual
orientations of the SRR slits in conventional metamaterials, either active or not,
the magnetic coupling is dominant.
These {\em magnetoinductive} systems support a new kind of waves with 
frequencies in a relatively narrow band of the optical type.
In the presence of nonlinearity, intrinsic localization in the form of discrete 
breathers (DBs) may occur generically by purely deterministic dynamics.
DBs are spatially localized and time-periodic excitations whose properties
have been extensively explored in the past \cite{Flach2008}; rigorous 
mathematical proofs of existence have been given for both energy conserved 
and dissipative lattices \cite{Mackay1994,Aubry1997}.
Moreover, they have been observed in a variety of physical systems including 
superconducting ones \cite{Binder2000,Trias2000}.
Dissipative DBs, in particular, may exist as a result of a power balance between 
input power and internal loss \cite{Marin2001}. Although the existence of 
dissipative DBs has been numerically demonstrated in both metallic SRR-based
\cite{Lazarides2006,Eleftheriou2008,Lazarides2008b,Eleftheriou2009,Tsironis2010} 
and SQUID-based metamaterials
\cite{Lazarides2008,Tsironis2009,Lazarides2012}, their experimental observation
is still lacking. In metallic metamaterials, losses constitute a major problem
that prevents breather formation; DB frequencies lie outside but close to the linear
frequency bands where high losses destroy self-focusing.
However, DBs could be in principle observed in SQUID-based metamaterials,
or in metamaterials where losses have been compensated by a gain mechanism. 
In ${\cal PT}$ metamaterials with alternating gain and loss, the net loss can become
in principle very low. Then, novel gain-driven DBs, whose existence has been also 
demonstrated numerically \cite{Lazarides2013,Tsironis2013}, could be also observed.

The present chapter focuses on the generation of stable or at least long-lived DBs in
dissipative-driven metallic and SQUID-based metamaterials, and novel gain-driven DBs in 
${\cal PT}$ metamaterials that rely on balanced gain and loss. In all cases, DBs result
from a power balance between intrinsic loss and input power. The input power comes 
either from an applied alternating magnetic field or, in the case of ${\cal PT}$ 
metamaterials, from an external source through the gain mechanism. For the sake of 
clarity in presentation, where temporal and spatial dependences are visible in single 
figure, we present only one-dimensional (1D) DBs. However, calculations with
the corresponding two-dimensional (2D) models reveal that these DBs are not destroyed by 
dimensionality, and moreover they may exist in the case of moderate anisotropy in the
coupling coefficients \cite{Eleftheriou2008,Lazarides2008}. In Sections 2 and 3,
the discrete model equations and dissipative DBs for metallic metamaterials and SQUID-based
metamaterials, respectively, are presented along with the corresponding frequency 
dispersions of the linearized systems. In Section 4, the model equations for a ${\cal PT}$
metamaterial with alternatingly gain and loss are presented in 1D, along
with the corresponding frequency dispersion. In this case, a condition for the 
metamaterial being in the exact ${\cal PT}$ phase is also obtained. Gain-driven DBs 
by either proper initialization or a purely dynamical mechanism are presented as well.
In Section 5 we conclude with a brief summary of the findings.

\section{Metalic SRR-Based Metamaterial}
\label{sec:1}
Consider a periodic arrangement of $N$ nonlinear, identical, metallic SRRs in 1D  
(Fig. 1), in two distinct configurations depending on the mutual orientation
of the SRRs in the array; the planar and the axial. 
Assuming that an SRR can be regarded as a resistive-inductive-capacitive $(RLC)$
oscillator featuring an Ohmic resistance $R$, self-inductance $L$, and capacitance 
$C$, its state can be described by the charge $Q$ in its capacitor and 
the current $I$ induced by an alternating magnetic field with appropriate polarization.
Assuming that the mutual orientations of the SRR 
slits are such that the magnetic interaction dominates over the electric one, the 
latter can be neglected. The magnetic coupling strength $\lambda$ can be quantified 
as the ratio of the mutual inductance $M$ between neighboring SRRs and the 
self-inductance of a single SRR, $L$, i.e., $\lambda={M}/L$.
Different configurations correspond to different signs in the coupling coefficients
between neighboring SRRs; thus $\lambda$ is negative (positive) between SRRs in the 
planar (axial) configuration. The most common configurations in 2D (not shown)
are the planar, where all SRR loops are in the same plane, or the planar-axial 
configuration where the SRRs have the planar configuration in one direction while
they have the axial configuration in the other direction 
\cite{Eleftheriou2008,Eleftheriou2009}. In 2D metamaterials on a square lattice 
there are two coupling coefficients $\lambda_{x}=M_x/L$ and $\lambda_{y}=M_y/L$, 
for coupling along the $x-$ and $y-$direction, respectively, with $M_x$ and $M_y$
being the corresponding mutual inductances.
The (normalized) dynamic equations for the state variables of each SRR in a 2D 
metamaterial read \cite{Shadrivov2006,Lazarides2006} 
\begin{figure}[h!]
\sidecaption[t]
\includegraphics[scale=.22]{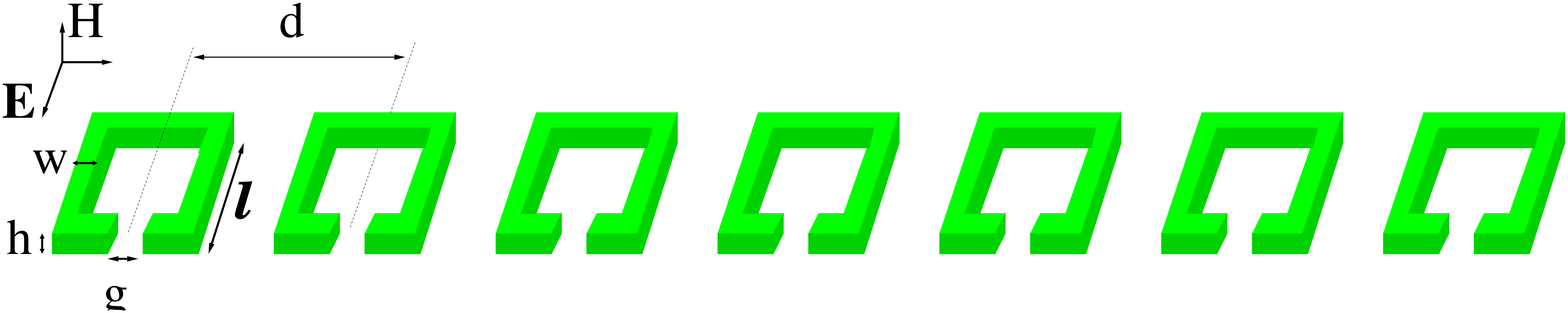}
\includegraphics[scale=.44,angle=-90]{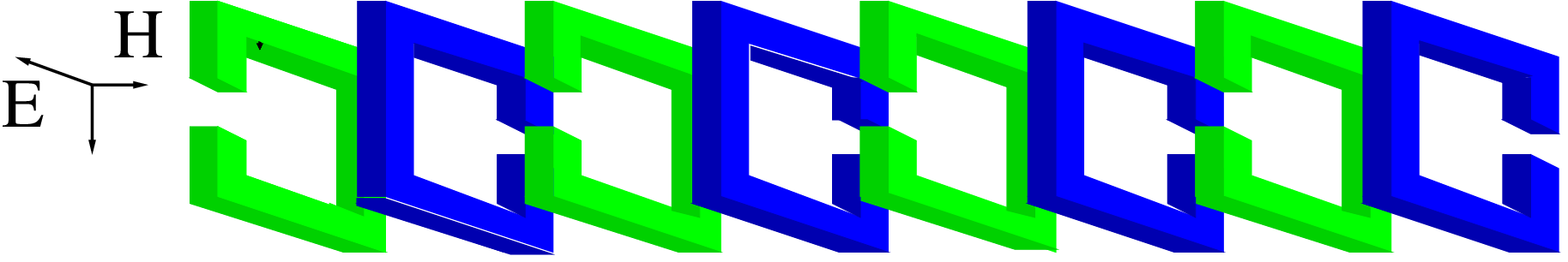}
\caption{
Schematic view of a one-dimensional array of split-ring resonators
in (upper) the planar geometry; (lower) the axial geometry.
The magnetic field is perpendicular to the planes of the rings.
}
\label{fig:1.01}       
\end{figure}
\begin{eqnarray}
\label{1.01}
  \frac{d^2}{d\tau^2}\left\{ q_{n,m} +\lambda_{x} \left( q_{n-1,m} +q_{n+1,m} \right) 
                                     +\lambda_{y} \left( q_{n,m-1} +q_{n,m+1} \right) \right\}
  \nonumber \\
   +\gamma \, \frac{d}{d\tau} q_{n,m} +f (q_{n,m}) 
   = \varepsilon_0 \, \sin(\Omega\tau) , 
\end{eqnarray}
where $q_{n,m}$ is the charge in the capacitor of the $(n.m)-$th SRR, 
$\tau$ is the normalized temporal variable,   
$\varepsilon_0$ is the amplitude of the induced emf, and 
$\gamma=R C_{\ell} \omega_{\ell}$ is the loss coefficient, with $C_{\ell}$ and 
$\omega_{\ell} =1/\sqrt{L \, C_{\ell}}$
being the linear capacitance and resonance frequency, respectively,
and $\Omega$ is the normalized driving frequency.
The derivative of the charge in the $(n,m)-$th SRR with respect to the temporal 
variable is the induced current in this SRR, i.e., $i_{n,m} =d q_{n,m} /d\tau$.
The function $f (q_{n,m})$ that provides the on-site nonlinearity, that may result from 
filling the SRR slits with a Kerr-type dielectric \cite{Zharov2003} or by mounting 
a diode into each SRR slit \cite{Lapine2003},
is approximated by $f(q_{n,m}) \simeq q_{n,m} -\chi q_{n,m}^{3}$,
where $\chi$ is a relatively small nonlinearity coefficient. 
The natural variables can be recovered from the normalized ones through the relations
\begin{eqnarray}
\label{1.02}
  t=\tau / \omega_\ell, \, \omega=\omega_\ell \Omega, \, Q_{n,m} = Q_c q_{n,m}, \, 
                          {\cal E}_0 =U_c \varepsilon_0, \, I_{n,m}=I_c i_{n,m} ,
\end{eqnarray}
with $I_c = U_c  \omega_\ell  C_\ell$, $Q_c=C_\ell  U_c$ and $U_c$ a characteristic
voltage. The frequency spectrum of linear excitations is obtained by substitution of
$q_{n,m}=A \cos(\kappa_x n + \kappa_y m -\Omega\tau)$ into Eqs. (\ref{1.01}) where
we also set $\chi=0$ and $\varepsilon_0 =0$. We thus obtain
\begin{eqnarray}
\label{1.03}
  \Omega_{\bf \kappa}  = 
  {[ 1 +2\, \lambda_{x} \, \cos(\kappa_x) 
       +2\, \lambda_{y} \, \cos(\kappa_y) ]}^{-1/2} ,
\end{eqnarray}
where ${\bf \kappa} =(\kappa_x, \kappa_y)$ is the normalized wavevector in 2D. 
\begin{figure}[h]
\sidecaption[t]
\includegraphics[scale=.38]{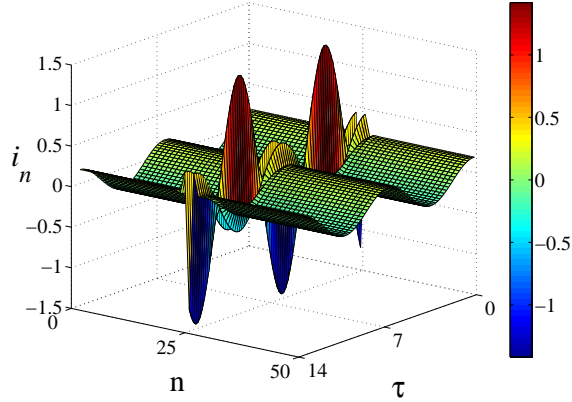}
\caption{
 Spatiotemporal evolution of a single-site, dissipative current breather,
 for $T_b=6.82$, $\lambda=-0.02$, $\gamma=0.01$, $\varepsilon_0=0.04$,
 $\chi=0.16$, and $N=50$. Both the background and the central
 breather site are oscillating with frequency $\Omega_b =2\pi/T_b$.
 The phase-difference of high and low current oscillations is almost $\pi$.
}
\label{fig:1.02}       
\end{figure}

Eqs. (\ref{1.01}) support dissipative DBs for relatively low losses, that can 
be generated with standard algorithms \cite{Marin2001,Martinez2003}.
Thus, we start by integrating the uncoupled equations while trying to adjust the 
driving frequency until we can identify two different, simultaneously stable solutions.
Let us denote the high and low amplitude solution with $q_h$ and $q_\ell$, 
respectively. Then, a trivial dissipative DB can be constructed by fixing the
amplitude of a particular SRR oscillator of the metamaterial to $q_h$  while the
amplitude of all the others is fixed to $q_\ell$. The corresponding derivatives with
respect to time, $d q_{n,m}/d\tau \equiv i_{n,m}$, are set to zero.
Using this trivial DB configuration as initial condition, the dynamic equations
are integrated while the coupling coefficients are switched on adiabatically.
It turns out that the trivial DB can be continued to nonzero couplings 
leading to dissipative DB formation
\cite{Lazarides2006,Eleftheriou2008,Eleftheriou2009,Tsironis2010}.
The spatiotemporal evolution of a typical, single-site dissipative DB in 1D is shown 
in Fig. 2 during approximately two periods of oscillation. Both the central
DB site and the background are oscillating with different amplitudes but
same frequency $\Omega_b=2\pi/T_b$, equal to that of the driver ($\Omega_b=\Omega$). 
Importantly,
high and low amplitude current oscillations occur in anti-phase, which indicates 
differences in response to the applied field that modify locally the magnetization 
\cite{Lazarides2006}. Depending on the frequency, DBs modify not only the magnitude 
but also the nature of the metamaterial response from paramagnetic to diamagnetic
or even extreme diamagnetic, the latter corresponding to negative magnetic 
permeability $\mu$.
Different types of DBs can be constructed using appropriate trivial breathers 
as initial conditions. A dissipative DB in the form of an oscillating
domain-wall that separates regions of a 1D metamaterial with different magnetizations,
is illustrated in Fig. 3.
Dissipative DBs may also be generated spontaneously in magnetic metamaterials with a 
binary configuration \cite{Molina2009,Lazarides2009,Lazarides2010a}, through a purely 
dynamical proccess that relies on the developement of modulational instability by a 
frequency-chirped driving field. This procedure is particularly well suited for DB 
generation in experimental situations, and has been applied successfully in 
micromechanical cantilever oscillator arrays \cite{Sato2003}.   
\begin{figure}[h]
\sidecaption[t]
\includegraphics[scale=.38]{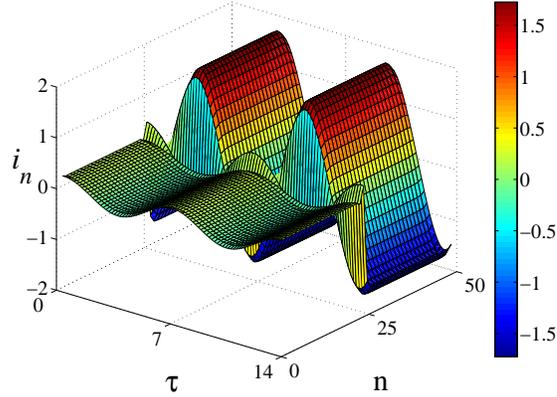}
\caption{
 Spatiotemporal evolution of a domain-wall dissipative breather,
 for $T_b=6.82$, $\lambda=-0.02$, $\gamma=0.01$, $\varepsilon_0=0.04$,
 $\chi=0.16$, and $N=50$.
 This peculiar type of breather
 separates regions of the metamaterial with different magnetizations.
}
\label{fig:1.03}       
\end{figure}

\section{rf SQUID Metamaterial}
\label{sec:2}
A SQUID metamaterial may be formed as a conventional metamaterial where the 
metallic elements have been replaced by rf SQUIDs \cite{Lazarides2007,Lazarides2008}.
The simplest rf SQUID, shown schematically in Fig. 4a, consists of a superconducting
path interrupted by a single JJ;
it constitutes the direct superconducting analogue of a nonlinear metallic SRR,
that plays the role of the 'magnetic atom' in SQUID-based metamaterials.
For a realistic description of a SQUID, the Resistively and Capacitively Shunted 
Junction (RCSJ) model is adopted \cite{Barone1982,Likharev1986}.
It results from shunting the ideal JJ, with critical current $I_c$, 
with a resistance $R$ and a capacitance $C$. The equivalent lumped circuit model 
for an rf SQUID in a magnetic field results from a series connection of the RCSJ
model for the JJ with an inductance $L$ and a flux source $\Phi_{ext}$ (Fig. 4b).
The dynamic equation for a single SQUID is then obtained by direct application of  
Kirkhhoff laws. 
SQUID metamaterials in 1D and 2D (square lattice) may be formed by repetition of the 
unit cell shown in Fig. 4a \cite{Kirtley2005}. 
Nonlinearity and discreteness, combined with weak coupling between neighboring SQUIDs
may lead in breather generation in this system as well 
\cite{Lazarides2008,Tsironis2009,Lazarides2012}.
The relevant dynamical variables in SQUID metamaterials are magnetic fluxes 
threading the SQUIDs, whose temporal evolution is described by the (normalized)  
equations 
\begin{eqnarray}
\label{2.01}
  \ddot{\phi}_{n,m} +\gamma \dot{\phi}_{n,m} +\phi_{n,m}
   +\beta\, \sin( 2 \pi \phi_{n,m} )
  -\lambda_x ( \phi_{n-1,m} +\phi_{n+1,m} ) 
\nonumber \\
  -\lambda_y ( \phi_{n,m-1} +\phi_{n,m+1} )
   = \phi_{eff} ,
\end{eqnarray}
where ${\phi}_{n,m}$ is the normalized flux threading the $(n,m)-$th SQUID,
$\lambda_x$ and $\lambda_y$ are the magnetic coupling coefficients between neighboring 
SQUIDs in the $x-$ and $y-$ direction, respectively,
the overdots denote differentiation with respect to the normalized time $\tau$,
$\beta =\frac{L I_c}{\Phi_0} =\frac{\beta_L}{2\pi}$ is the SQUID parameter,
and
$\gamma =\frac{1}{R} \sqrt{\frac{L}{C}}$ is the loss coefficient of each individual 
SQUID, with $\Phi_0$ being the magnetic flux quantum. The coupling coefficients
are defined as in the previous Section, i.e., $\lambda_x =M_x/L$ and $\lambda_y =M_y/L$
in the $x-$ and $y-$direction, respectively, with $M_x$ and $M_y$ being the mutual
inductances. SQUID arrays are fabricated in the planar geometry, and therefore 
the values of the $M_x$ and $M_y$ are negative.
The rf SQUID exhibits strong resonant response to an alternating magnetic field at
a particular frequency $\omega_{SQ} = \omega_{LC} \sqrt{ 1 +\beta_L }$,
with $\omega_{LC} =1 / \sqrt{L C}$ being its corresponding inductive-capacitive 
frequency. 
\begin{figure}[h]
\sidecaption[t]
\includegraphics[scale=.40]{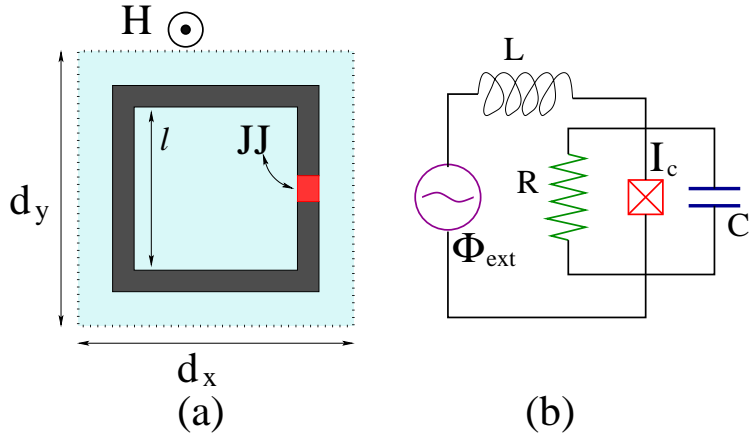}
\caption{
(a) Schematic drawing of the unit cell of a SQUID metamaterial with one SQUID per
cell. The applied magnetic field ${\bf H}(t)$ is perpendicular to the SQUID plane. 
(b) The equivalent Resistively and Capacitively Shunted Junction ($RCSJ$) model for a 
an rf SQUID with one Josephson junction driven by a flux source $\Phi_{ext}$.}
\label{fig:4}       
\end{figure}
In equations (\ref{2.01}), the fluxes are normalized to $\Phi_0$,
while the temporal variable is normalized to $\omega_{LC}^{-1}$. 
The term on the right-hand-side of Eqs. (\ref{2.01}) is the {\em effective}
external flux $\phi_{eff} =[1-2(\lambda_x +\lambda_y)] \phi_{ext}$, where  
$\phi_{ext} =\phi_{dc}  +\phi_{ac} \cos(\Omega \tau )$ is the flux due to the applied,
spacially uniform magnetic field.
The latter may have both constant (dc) and alternating (ac) terms,
resulting from a constant and an alternating magnetic field with normalized 
frequency $\Omega$, respectively. 
A dc field may be used to provide tunability of the SQUID resonance over relatively
wide frequency intervals \cite{Poletto2009,Jung2013}. The resonance shifts due to an
applied dc and/or ac fields may also reproduced numerically from the SQUID 
model equation \cite{Lazarides2008,Lazarides2012}.

By linearization of the free (i.e., $\gamma=0$, $\phi_{ext} =0$) equations (\ref{2.01})
and substitution of the trial solution 
$\phi= A\, \exp[i (\kappa_x n + \kappa_y m - \Omega_{\bf \kappa} \tau)]$, we obtain
\begin{eqnarray}
  \label{2.04}
   \Omega_{\bf \kappa} = \sqrt{1 + \beta_L -2( \lambda_x \, \cos \kappa_x
                                +\lambda_y \, \cos \kappa_y ) } ,
\end{eqnarray}
where $\Omega_{\bf \kappa}$ is the eigenfrequency at wavevector
${\bf \kappa} =(\kappa_{x},\kappa_{y}) =(k_x d_x^{-1},k_y d_y^{-1})$,
with $k_x$, $k_y$ being the wavevector components in natural units and 
$d_{x}$, $d_{y}$ the center-to-center distance between neighboring SQUIDs in the $x-$
and $y-$direction, respectively. 
Eq. (\ref{2.04}) provides the linear frequency dispersion of magnetoinductive
flux-waves whose typical form is shown in Fig. 5 \cite{Lazarides2008}, that is very 
similar to that of metallic metamaterials in 2D \cite{Shamonina2002}.
In the absence of losses ($\gamma=0$) and ac flux $\phi_{ac}=0$, Eqs. (\ref{2.01})
can be obtained from the Hamiltonian 
\begin{eqnarray}
\label{2.05}
   \frac{H}{E_J} =  \sum_{n,m} \left\{ \frac{\pi}{\beta} 
             \dot{\phi}_{n,m}^2 + u_{n,m} \right\} 
    \nonumber \\
    -\frac{2\pi}{\beta} \sum_{n,m}  \left\{
         \lambda_x (\phi_{n,m} -\phi_{dc} ) (\phi_{n-1,m} -\phi_{dc} )
        +\lambda_y (\phi_{n,m} -\phi_{dc} ) (\phi_{n,m-1} -\phi_{dc} ) \right\} ,
\end{eqnarray}
where $E_j = I_c \Phi_0 / (2\pi)$ is the Josephson energy, and 
\begin{eqnarray}
\label{2.06}
   u_{n,m} =\frac{\pi}{\beta} (\phi_{n,m} -\phi_{dc})^2 -\cos(2\pi \phi_{n,m}) ,
\end{eqnarray}
is the on-site potential. The flux-balance relation for the 2D SQUID
metamaterial, expressed in the same order of approximation as the dynamic equations, 
reads
\begin{eqnarray}
\label{2.07}
   \phi_{n,m}^{loc} = \phi_{eff} +\beta \, i_{n,m} ,
\end{eqnarray}
where  
$\phi_{n,m}^{loc} = \phi_{n,m} -\lambda_x (\phi_{n-1,m} +\phi_{n+1,m})
              -\lambda_x (\phi_{n,m-1} +\phi_{n,m+1})$,
is the local flux at the lattice site $(n,m)$. Eq. (\ref{2.07})
generalizes the corresponding flux-balance relation of a single SQUID.
\begin{figure}[h!]
\sidecaption[t]
\includegraphics[scale=.58]{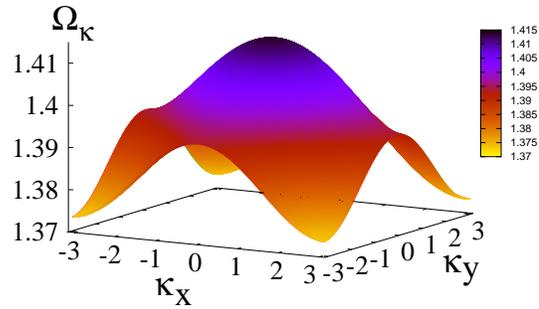}
\caption{
The linear frequency dispersion $\Omega_{\vec\kappa}$ plotted as a function of $\kappa_x$
and $\kappa_y$ for a two-dimensional SQUID metamaterial with $\lambda_x = \lambda_y =-0.014$
and $\beta=0.15$. The band extends from $\Omega_{min}=1.374$ to $\Omega_{min}=1.414$.
}
\label{fig:5}       
\end{figure}

For generating dissipative DBs in SQUID metamaterials we use two different 
approaches; first, we employ the same algorithm as in the previous Section, and
second, by introducing weak disorder in the SQUID parameter $\beta$.
Estimates for the coupling strength between SQUIDs obtained using data 
from the literature \cite{Kirtley2005}, give $|\lambda_{x,y}| \simeq 0.014$
for a SQUID metamaterial with isotropic coupling.
These values are very reasonable and consistent with our weak coupling approximation.
They are also of the same order of magnitude with the corresponding ones obtained for 
metallic metamaterials. However, we sometimes use higher values of the coupling coefficients
in order to demonstrate that breather generation is not just a marginal effect. 
The SQUID potential $u_{n,m}$ given in Eq. (\ref{2.06}) allows for many more 
possibilities for breather generation. The number and the location of minima of 
$u_{n,m}$ can be controlled either by the parameter $\beta$ or in real-time by 
a dc applied flux $\phi_{dc}$. While for $\beta_L <1$ there is only one minimum,
multiple minima appear for $\beta_L >1$ with their number increasing with further
increasing $\beta_L$. 
The dc flux, on the other hand, may both create new minima and move their 
positions to different flux values. For example, for $\phi_{dc}=0.5$ ($\beta_L <1$)
the potential takes the form of a symmetric double-well. Then, the construction
of trivial breather states is a rather obvious task; 
one may choose flux states with high and low flux amplitude corresponding to 
the two minima of the potential. Then, one of the SQUIDs is set to the high 
amplitude state and the other ones to the low amplitude states. A typical 
dissipative DB in a multistable potential in 1D is shown in Fig. 7.
This type of breather cannot appear in metallic metamaterials, for which the 
on-site potential has a single minimum. The temporal evolution of the DB diverges 
significantly from a sinusoidal, due to strong nonlinearities even at low powers;
this is another peculiarity resulting from the form of $u_{n,m}$.
In this case, all SQUIDs oscillate almost in phase, while they differ only in their 
current oscillation amplitude. 
\begin{figure}[h!]
\sidecaption[t]
\includegraphics[scale=.37]{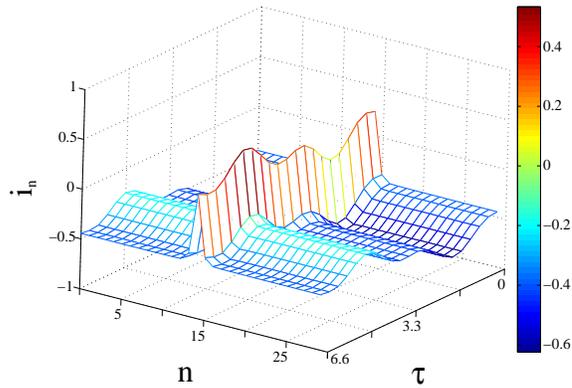}
\caption{
Spatiotemporal evolution of a single-site dissipative breather during one period 
of oscillation, for a SQUID metamaterial with
$\beta=1.27$, $\gamma=0.001$, $N=30$, $\lambda=-0.1$, and $T_b=6.6$,
$\phi_{dc}=0.5$, and $\phi_{ac}=0.2$.
Note the phase-coherence and the non-sinusoidal time-dependence of the oscillations. 
}
\label{fig:6}       
\end{figure}

The strong nonlinearity in the SQUIDs manifests itself also with the existence of 
several simultaneously stable solutions. The multistability of SQUID states
implies multistability for possible DB configurations; indeed, by combination
of two or more simultaneously stable single SQUID states for the construction
of 'trivial breathers', we may generate simultaneously stable DBs \cite{Lazarides2008}.
Typical DBs of this type in 1D look like that in Fig. 7, which exhibits features similar
to those of the corresponding DBs in metallic metamaterials (see e.g. Fig. 2).  
This type of DB may change locally the nature of the magnetic response from diamagnetic
to paramagnetic (or vice versa) just as in metallic metamaterials.
\begin{figure}[t]
\sidecaption[t]
\includegraphics[scale=.37]{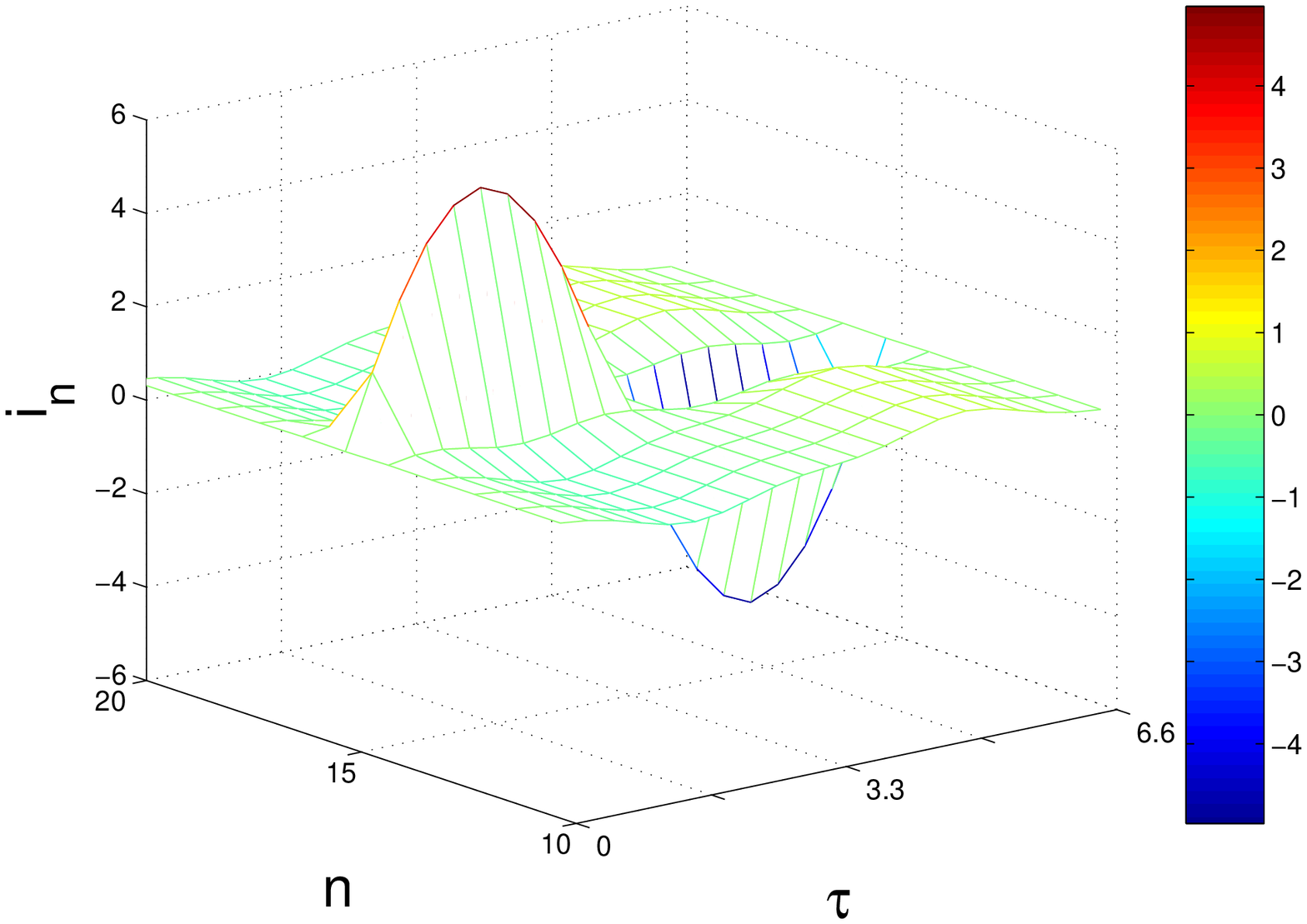}
\caption{
Spatiotemporal evolution of single-site dissipative breather during one period 
of oscillation, for a SQUID metamaterial with
$\beta=1.27$, $\gamma=0.001$, $N=30$, $\lambda=0.1$, and $T_b=6.6$,
$\phi_{dc}=0$, and $\phi_{ac}=0.6$.
}
\label{fig:7}       
\end{figure}
\begin{figure}[t]
\sidecaption[t]
\includegraphics[scale=.37]{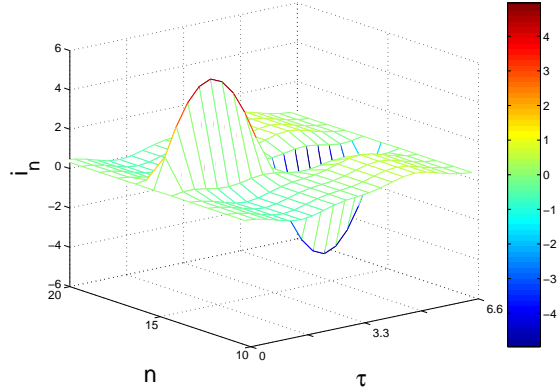}
\caption{
Spatiotemporal evolution of a dissipative, period-3 breather during three
driver periods, for a SQUID metamaterial with $N=30$, $\phi_{dc}=0$, $\phi_{ac}=1.2$, 
$\gamma=0.001$, $\beta=1.27$, and $T_b=12.57$.
}
\label{fig:8}       
\end{figure}
In most cases the DB frequency equals to that of the driver. However, there
is also the possibility for multiperiodic DBs to appear, whose period $T_b$
is an integer multiple of that of the driver $T=2\pi/\Omega$. A period-3 
dissipative DB, with $T_b =3 T$, is shown in Fig. 8 \cite{Lazarides2008}.
Poincar\'e diagrams for the trajectories of the central DB site against those 
for the sites in the background (not shown) confirm the observed multiperiodicity.
Although we have presented mostly single-site and "bright" DBs, 
multi-site as well as 'dark' DBs can be also generated by appropriate choice
of a trivial breather \cite{Eleftheriou2008,Tsironis2010,Lazarides2012}.
The linear stability of dissipative DBs can be addressed through the eigenvalues
of the Floquet matrix (Floquet multipliers). A dissipative DB is linearly stable
when all its Floquet multipliers lie on a circle of radius $R_e = \exp(-\gamma T_b/2)$
in the complex plane \cite{Marin2001}. The breathers presented here have been found 
to be linearly stable. Moreover, they were let to evolve for long times 
(i.e., more than $10^5~T_b$) without any observable change in their shapes.
\begin{figure}[h]
\sidecaption[t]
\includegraphics[scale=.45]{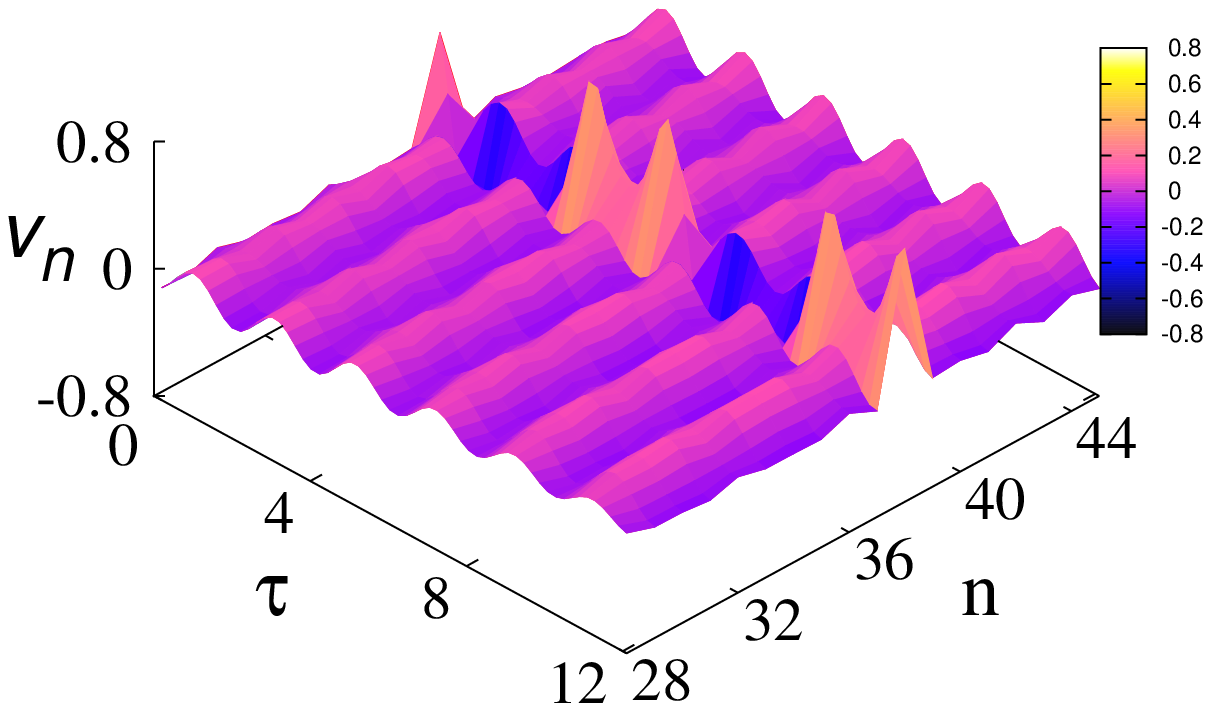}
\includegraphics[scale=.45]{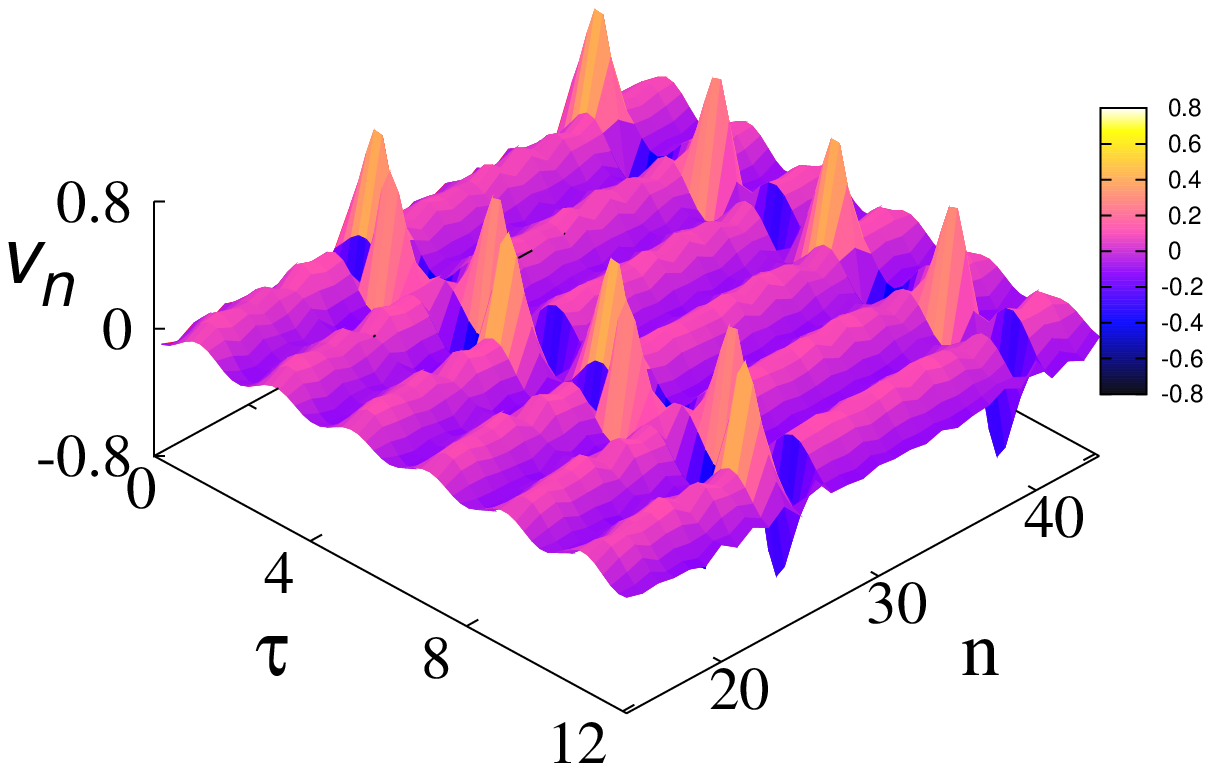}
\caption{
Spatiotemporal evolution of dissipative breathers excited spontaneously in
weakly disordered rf SQUID metamaterials in 1D during six periods.
The voltage amplitude in the Josephson junction of the $n$th SQUID $v_n=d\phi/d\tau$
is plotted on the $\tau - n$ plane
for $\phi_{dc}=0$, $\phi_{ac}=0.03$, $\beta=1.27$, $\gamma=0.001$, $\lambda=-0.0014$,
$\omega=3.11$, and $N=50$. The left and right panels correspond to different
configurations of disorder.
}
\label{fig:9}       
\end{figure}

The algorithm used above requires initialization of the system with a particular 
configuration ('trivial breather'), which is may not be always convinient in 
experimental situations.
However, in SQUID metamaterials spontaneous DB generation may be 
facilitated by the existence of weak disorder due to limited accuracy during
fabrication. 
In a particular realization of an rf SQUID metamaterial, the elements cannot be
completely identical but the values of their parameters fluctuate around a mean.
The critical current $I_c$ of the JJs seems to be more sensitive to misperfections
in fabrication, since it depends exponentially on the thickness of the insulating 
dielectric. Then, fluctuations of $I_c$ result in fluctuations in the SQUID parameters 
$\beta$, which in turn determine the SQUID frequency.
We have performed numerical calculations for a SQUID metamaterial in 1D with $\beta$
allowed to vary randomly within $\pm 1\%$ around its mean value.
Then, by integrating Eqs. (\ref{2.01}) for a number different configurations
of disorder we obtained in most cases spontaneously generated dissipative DBs.
For this approach to work, it is required that the coupling between SQUIDs is very weak.
Typical results for the spatiotemporal evolution of spontaneously generated 
dissipative DBs in disordered SQUID metamaterials are shown in Fig. 9, 
where the oscillations of the instantaneous voltage $v_{n} = d\phi_{n} /dt$ are plotted
on the $n - \tau$ plane.
The left and right panels correspond to two different configurations of disorder,
while all the other parameters are fixed. 
It is observed that the number of generated DBs in the two cases is different
(one and three, respectively) with the DB central sites located at different positions. 
In the left panel, the period of voltage oscillation in the central DB sites is twice 
that of the driver, so that it is actually a period-2 breather.

\section{${\cal PT}-$Symmetric Metamaterial}
\label{sec:3}
Consider a 1D array of dimers, each comprising two nonlinear SRRs;
one with loss and the other with equal amount of gain (Fig. 10). The SRRs may be 
arranged in a chain with two different configurations. As shown schematically in 
Fig. 10, the SRRs in the array may be either equidistant (Fig. 10a) or the distance 
between them may be modulated according to a binary pattern (Fig.10b). 
Due to balanced gain and loss in each dimer, these configurations obey a combined 
${\cal PT}-$symmetry. Building ${\cal PT}-$symmetric metamaterials may provide a 
way to overcome losses and moreover to reveal new extraordinary properties. 
These systems undergo spontaneous symmetry breaking from the exact 
${\cal PT}$ phase, where all eigenfrequencies are real, to the broken ${\cal PT}$ 
phase, where at least one pair of eigenfrequencies are complex, with the variation
of a control (gain/loss) parameter. For low values of the gain/loss parameter, 
${\cal PT}-$symmetric systems are usually in the exact phase; 
however, when that parameter exceeds a critical value, the system goes into the 
broken phase. For linear ${\cal PT}-$symmetric systems, stable solutions exist 
only in the exact phase. In this Section we obtain the linear frequency spectrum of 
a linear ${\cal PT}$ metamaterial, and the conditions for having stable solutions.
We then demonstrate numerically DB generation in a nonlinear 
${\cal PT}$ metamaterial model in the dimer chain configuration. 
Specifically, it is demonstrated 
that long-lived DBs that are powered by the gain mechanism can be generated either
by proper initialization of the ${\cal PT}$ metamaterial or purely dynamically
through external driving. 
\begin{figure}[h!]
\sidecaption[t]
\includegraphics[scale=.22]{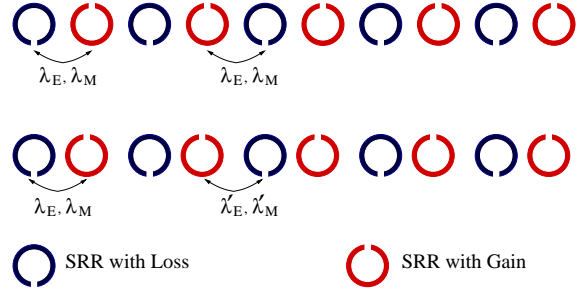}
\caption{
Schematic of a ${\cal PT}$ metamaterial.
Upper panel: all the SRRs are equidistant.
Lower panel: the separation between SRRs is modulated according to a binary pattern
(${\cal PT}$ dimer chain).
}
\label{fig:10}       
\end{figure}

In the equivalent circuit model picture
\cite{Shadrivov2006,Lazarides2006,Molina2009,Lazarides2011,Rosanov2011},
extended for the ${\cal PT}$ dimer chain, the dynamics of the charge $q_n$
in the capacitor of the $n$th SRR is governed by
\begin{eqnarray}
\label{3.01}
   \lambda_M' \ddot{q}_{2n} +\ddot{q}_{2n+1} +\lambda_M \ddot{q}_{2n+2}
  +\lambda_E' q_{2n} +q_{2n+1} +\lambda_E q_{2n+2}
  +\gamma \dot{q}_{2n+1} 
 \nonumber \\
  +\alpha q_{2n+1}^2 +\beta q_{2n+1}^3 
   =\varepsilon_0 \sin(\Omega \tau) \\ 
\label{3.02}
   \lambda_M \ddot{q}_{2n-1} +\ddot{q}_{2n} +\lambda_M' \ddot{q}_{2n+1}
   +\lambda_E q_{2n-1} +q_{2n} +\lambda_E' q_{2n+1} 
   -\gamma \dot{q}_{2n} 
 \nonumber \\
   +\alpha {q}_{2n}^2 +\beta {q}_{2n}^3
    =\varepsilon_0 \sin(\Omega \tau) ,
\end{eqnarray}
where
$\lambda_M, \lambda_M'$ and $\lambda_E, \lambda_E'$ are the magnetic and electric
interaction coefficients, respectively, between nearest neighbors,
$\alpha$ and $\beta$ are nonlinear coefficients,
$\gamma$ is the gain/loss coefficient ($\gamma >0$),
$\varepsilon_0$ is the amplitude of the external driving voltage,
while $\Omega$ and $\tau$ are the driving frequency and temporal variable,
respectively, normalized to $\omega_0 =1/\sqrt{L C_0}$ and $\omega_0^{-1}$,
respectively, with $C_0$ being the linear capacitance.
The total number of SRRs is an even integer $N$, so that there are $N/2$ 
${\cal PT}$ symmetric dimers.
In the following, we consider that the relative orientation of the SRRs in the 
chain is such that the magnetic coupling dominates, while the electric coupling
can be neglected ($\lambda_E = \lambda_E' =0$) \cite{Hesmer2007}.
\begin{figure}[h]
\sidecaption[t]
\includegraphics[scale=.48]{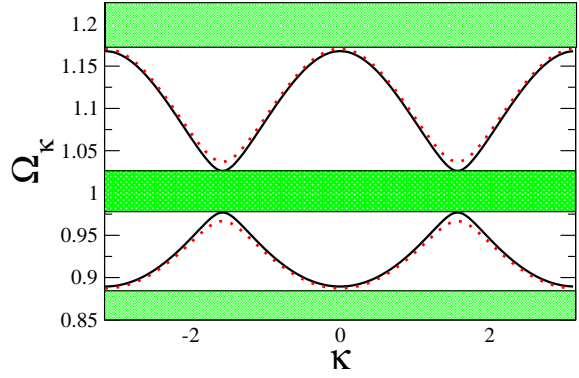}
\caption{
Frequency bands for a ${\cal PT}-$symmetric dimer chain with balanced
gain and loss for $\lambda_M=-0.17$, $\lambda_M' =-0.10$, and
$\gamma =0.05$ (black solid curves); $\gamma =0$ (red dotted curves).
The forbidden frequency regions are indicated in green (dark) color.
Note that the gain/loss coefficient $\gamma$ has a minor effect on the 
dispersion curves.
}
\label{fig:11}       
\end{figure}

In the linear regime, without external driving, we set $\alpha=\beta=0$ and 
$\varepsilon_0 =0$ in Eqs. (\ref{3.01}) and (\ref{3.02}). We keep however the gain/loss 
terms that are proportional to $\pm \gamma$ and provide ${\cal PT}-$symmetry.
We then substitute 
\begin{eqnarray}
\label{3.03}
  q_{2n} = A \exp[i( 2 n \kappa -\Omega_\kappa \tau)], \qquad
  q_{2n+1}=B \exp[i( (2 n+1) \kappa -\Omega_\kappa \tau)] ,
\end{eqnarray}
where $\kappa$ is the normalized wavevector, and request nontrivial solutions for the
resulting stationary problem. We thus obtain the frequency dispersion
\begin{eqnarray}
 \label{3.08}
  \Omega_\kappa^2 = \frac{2-\gamma^2 \pm \sqrt{\gamma^4 -2 \gamma^2 
               +(\lambda_M -\lambda_M')^2 +\mu_\kappa \mu_\kappa'} }
             {2 (1 -(\lambda_M -\lambda_M')^2 -\mu_\kappa \mu_\kappa')} ,
\end{eqnarray}
where
$\mu_\kappa = 2 \lambda_M \cos(\kappa)$,
$\mu_\kappa' = 2 \lambda_M' \cos(\kappa)$.
The condition for having real $\Omega_\kappa$ for any $\kappa$ in the earlier equation
then reads
\begin{eqnarray}
\label{3.09}
  \cos^2 (\kappa) \geq \frac{\gamma^2 (2 -\gamma^2) -(\lambda_M -\lambda_M')^2}
                            {4 \lambda_M \lambda_M'} ,
\end{eqnarray}
From Eq. (\ref{3.09}) it is easy to see that for $\lambda_M =\lambda_M'$,
corresponding to the equidistant SRR configuration, the condition for real $\Omega_\kappa$
for all $\kappa$ cannot be satisfied for any positive value of the gain/loss 
coefficient $\gamma$. This result implies that a large ${\cal PT}-$symmetric SRR array 
cannot be in the exact phase and therefore stable, stationary solutions cannot exist.
To the contrary, for $\lambda_M \neq \lambda_M'$, i.e., for a ${\cal PT}$ dimer chain,
the condition (\ref{3.09}) is satisfied for all $\kappa$'s for 
$\gamma \leq \gamma_c \simeq |\lambda_M -\lambda_M'|$,  ($\gamma^4 \simeq 0$).
In the exact phase ($\gamma < \gamma_c$), the ${\cal PT}-$symmetric dimer array has
a gapped spectrum with two frequency bands, as shown in Fig. 11.
\begin{figure}[h!]
\sidecaption[t]
\includegraphics[scale=.74]{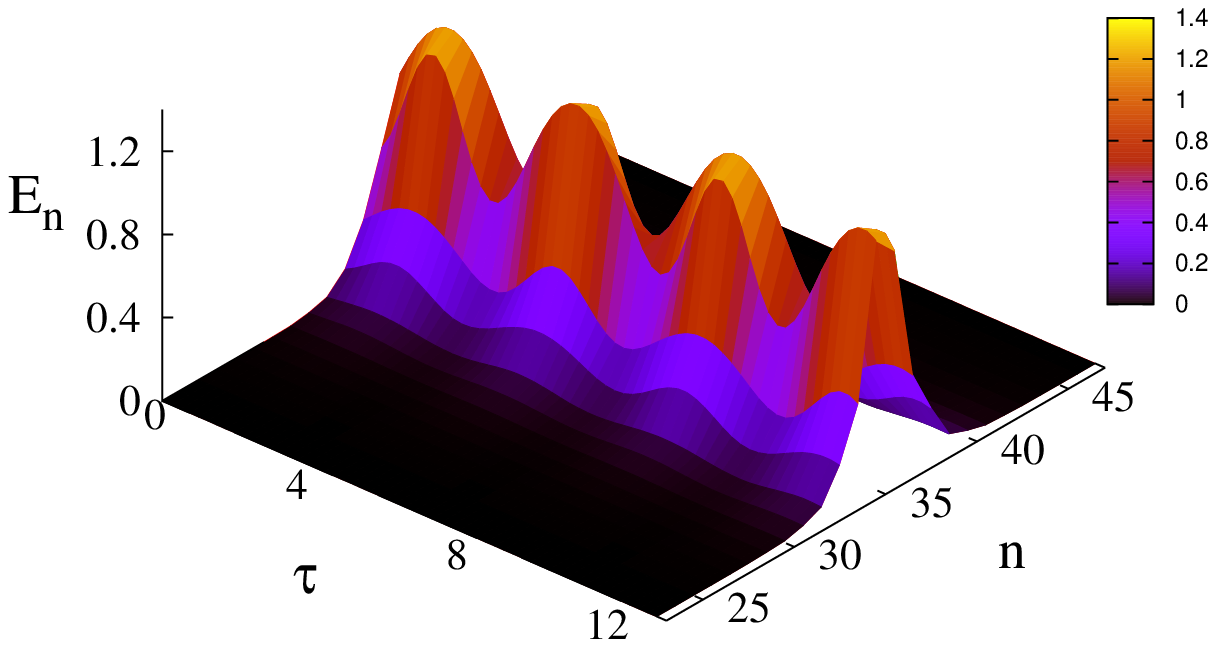}
\includegraphics[scale=.74]{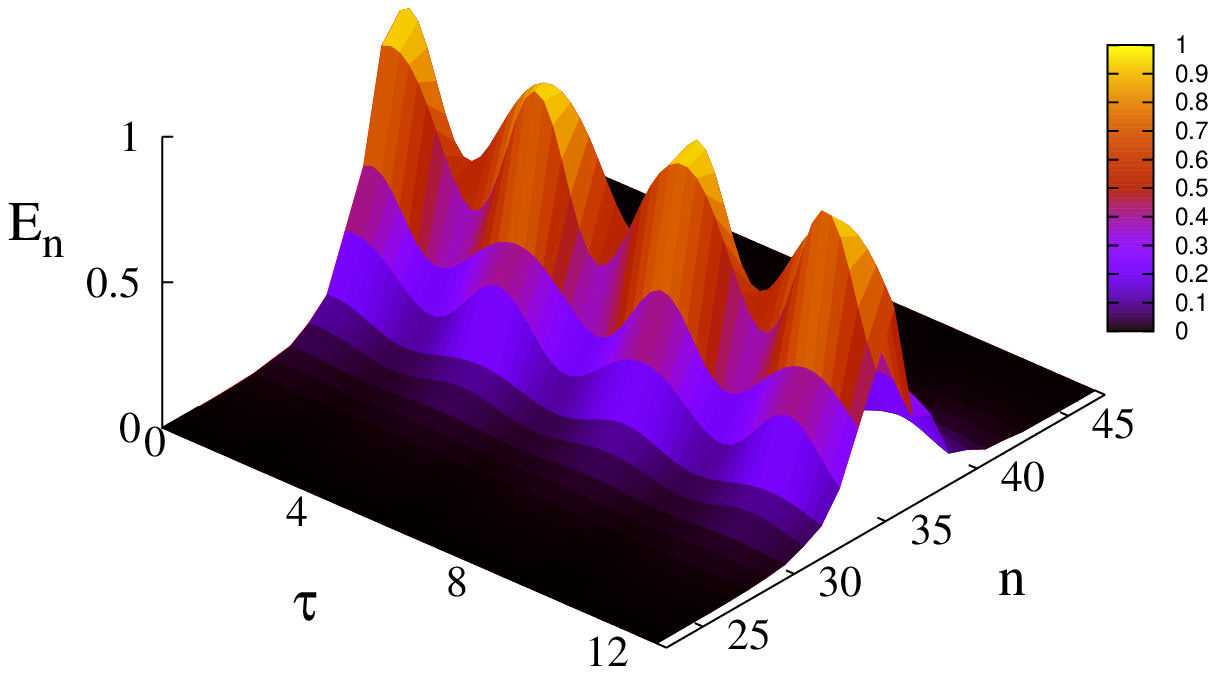}
\caption{
Spatiotemporal evolution of the energy density $E_n$ on the $n-\tau$ plane during 
two periods of oscillation for a ${\cal PT}$ metamaterial with $N=70$, $N_\ell =10$,
$\gamma=0.002$, $\lambda_M' =-0.10$, $\lambda_E =\lambda_E' =0$, and
(upper) $\lambda_M=-0.17$; (lower) $\lambda_M=-0.21$.
Energy localization at two neighboring sites, one with gain and one with loss,
is clearly observed. 
}
\label{fig:12}       
\end{figure}

Eqs. (\ref{3.01}) and (\ref{3.02}), implemented with the boundary conditions 
$q_0 (\tau) = q_{N+1} (\tau) =0$, are integrated numerically with
\begin{equation}
\label{3.10}
  q_m (0) =(-1)^{m-1} {\rm sech}(m/2) , \qquad \dot{q}_m (0) =0 ,
\end{equation}
and $\varepsilon_0 =0$. The nonlinear coefficients are fixed to $\alpha=-0.4$ and
$\beta=0.08$, values that are typical for a diode \cite{Lazarides2011},
while $\gamma$ is chosen so that the ${\cal PT}$ metamaterial is well into the exact
phase. The coupling coefficients are chosen relatively
large in comparison with the values reported in the literature for clarity.
However, breathers appear generically even for much lower coupling values. 
In order to prevent instabilities that would result in divergence of the energy
at particular sites in finite time scales, 
we embbed the ${\cal PT}-$symmetric dimer chain into a lossy dimer chain.
In practice, we consider a longer dimer chain with total number of SRRs $N+2N_\ell$;
then we replace the gain with equivalent amount of loss at exactly $N_\ell$
SRRs at each end of the extended chain. That helps the excess energy to go
smoothly away during the long transient phase of integration, living behind stable 
(or at least very long-lived, for more than $\sim 10^8$ time units) DBs
\cite{Lazarides2013,Tsironis2013}. 
Typical energy density, $E_n$, plots in the $n-\tau$ plane
are shown in Fig. 12. A large amount of the total energy $E_{tot}=\sum_n E_n$
is concentrated into two neighboring sites (SRRs) that belong to the same dimer.
Thus, the fundamental breather excitation in the ${\cal PT}$ metamaterial
is actually a two-site DB, and not a single-site DB like those presented in the 
previous Sections.
The energy densities also exhibit regular oscillations, as it is expected for 
${\cal PT}-$symmetric systems. Inspection of the corresponding instantaneous current 
profiles (Fig. 13) $i_n$ as a function of $n$, reveal that these DBs are 
neither symmetric nor antisymmetric at the single SRR level.
\begin{figure}[h]
\sidecaption[t]
\includegraphics[scale=.45]{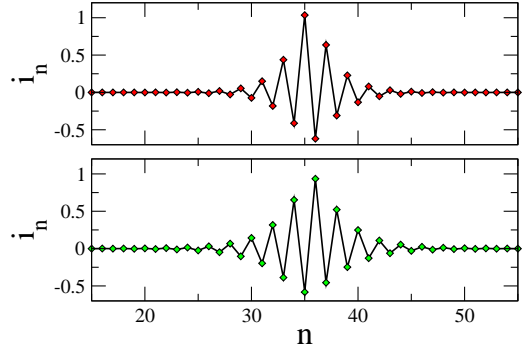}
\caption{
Gain-driven, current breather profiles $i_n$ as a function of $n$ at maximum 
amplitude, for the breathers shown in Fig. 12. These profiles are neither symmetric 
nor antisymmetric at the single SRR level.
}
\label{fig:13}       
\end{figure}

For a gapped linear spectrum, large amplitude linear modes become unstable
in the presence of driving and nonlinearity. If the curvature of the dispersion
curve in the region of such a mode is positive and the lattice potential is
soft, large amplitude modes become unstable with respect to formation of DBs
in the gap below the linear spectrum \cite{Sato2003}.
For the parameters used in Fig. 11, the bottom of the lower band is located
at $\Omega_0 = 2\pi/T_0 \simeq 0.887$, where the curvature is positive.
Moreover, the SRRs are subjected to soft on-site potentials for the selected
values of $\alpha$ and $\beta$.
Then, DBs can be generated spontaneously by a frequency chirped alternating
driver; after the driver is turned off, the breathers are driven solely by gain.
A similar procedure has been applied succesfuly to lossy nonlinear metamaterials
with a binary structure \cite{Molina2009,Lazarides2009,Lazarides2010a}.
Gain-driven DBs that are spontaneously generated by a frequency chirped
driver can be visualized on an energy density map on the $n-\tau$ plane (Fig. 14).
We use the following procedure:
\begin{figure}[h!]
\sidecaption[t]
\includegraphics[angle=0, width=0.62 \linewidth]{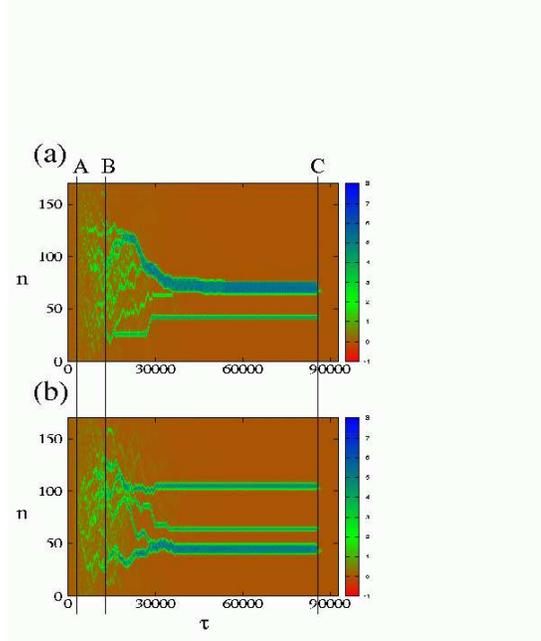}
\caption{
The energy density, $E_n$, mapped on the $n-\tau$ plane for a ${\cal PT}$
metamaterial (dimer chain) with $N=170$, $N_\ell =10$, $\Omega_0 =0.887$, 
$\gamma=0.002$,
$\lambda_M =-0.17$, $\lambda_M' =-0.10$ ($\lambda_E =\lambda_E' =0$), and
(a) $\varepsilon_0 =0.085$; (b) $\varepsilon_0 =0.095$.
The vertical lines separate different stages in the chirping procedure.
Between the points B and C, the breathers indicated by the blue-green (dark)
horizontal segments are solely driven by the gain.
They show a clear tendency to merge together forming wide multi-site structures.  
}
\label{fig:14}       
\end{figure}

\noindent $\bullet$
At time $\tau=0$, we start integrating Eqs. (\ref{3.01}) and (\ref{3.02}) with zero
initial state without external driving for $500~T_0 \simeq 3500$ time units (t.u.),
to allow for significant development of large amplitude modes.

\noindent $\bullet$
At time $\tau \simeq 3500$ t.u. (point A on Fig. 14), the driver is switched-on
with low-amplitude and frequency slightly above $\Omega_0$
($1.01~\Omega_0 \simeq 0.894$). The frequency is then chirped downwards with
time to induce instability for the next $10,600$ t.u. ($\sim 1500~T_0$),
until it is well below $\Omega_0$ ($0.997~\Omega_0 \simeq 0.882$).
During that phase, a large number of excitations are generated that move and
strongly interact to each other, eventually merging into a small number of high
amplitude multi-breathers.

\noindent $\bullet$
At time $\tau \simeq 14,100$ t.u. (point B on Fig. 14), the driver is switched off
and the DBs are solely driven by the gain (gain-driven phase). 
They continue to interact to each other until they reach an apparently stationary 
state. 
The high density horizontal segments between points B and C in Fig. 14 present 
precisely those stationary gain-driven (multi-)breathers generated through
the dynamics.

\noindent $\bullet$
At time $\tau \sim 85150$ t.u. (point C on Fig. 14), the gain is replaced by equal
amount of loss, and the breathers die out rapidly.

The above procedure is very sensitive to parameter variations of the external fields;
the number of DBs as well as their locations in the lattice may change with slight
parameter variation (Fig. 14). The DBs formed during the chirping phase continue
to interact with each other for longer times, showing a tendency to merge together 
into wide multi-site structures that occupy an even number of sites.
The frequency $\Omega_b$ of these DBs lies slightly below the lower band of the 
linear spectrum. Gain-driven DBs may still be generated by the above procedure
when there is a slight imbalance between gain and loss \cite{Tsironis2013}. 
The gain/loss imbalance is 
manifested either as a decay or growth of the total energy, in a timescale that
depends on the amount of imbalance. When loss exceeds gain, a multibreather gradually
looses its energy, since its excited sites at its end-points fall the one after the 
other in a low amplitude state. In the opposite case, where gain exceeds loss,
a multibreather slowly gains energy and becomes wider. Thus, in a realistic
experimental situation where gain/loss balance is only approximate,
it would be still possible for breathers to be observed at relatively short time-scales.

\section{Summary}
Breather excitations appear generically in nonlinear metallic, SQUID-based, and 
${\cal PT}$ metamaterials in the presence of dissipation that is always present in
practice. These {\em dissipative breathers} can be accurately
constructed either by using standard algorithms that require a proper initialization 
of the system or by dynamic effects that
are more suitable in real experimental situations. Low losses, a prerequisite for 
DB observation, can be achieved either by inserting electronic elements that 
provide gain, or by replacing the metallic SRRs with superconducting ones.
In conventional, metallic metamaterials and ${\cal PT}$ metamaterials, which expose 
their unusual properties when driven by an alternating field, DB generation
by frequency chirping seems to be a convinient approach well suited for experiments.
Also, as it is demonstrated for SQUID-based metamaterials, the presence of weak
disorder may trigger intrinsic localization that is subsequently evolved into a breather
through self-focusing. Dissipative breathers are certainly closer to reality than 
their energy-conserving counterparts (i.e., Hamiltonian breathers) and result 
from a power-balance of intrinsic losses due to dissipation and input power.
The latter may either come from an externally applied alternating field, as in the 
case of metallic and SQUID-based metamaterials, or by a particular gain mechanism,
as in the case of the proposed ${\cal PT}$ metamaterials 
\cite{Lazarides2013,Tsironis2013}.   
Dissipative breathers are very robust since they correspond to attractors of the
"motion" in a high-dimensional phase space, and relatively weak perturbations 
disappear in short time-scales. Moreover, they exhibit features not seen in
Hamiltonian breathers; e.g., current oscillations appear in all the elements of 
a metamaterial in a dissipative breather configuration. High and low current 
oscillations are almost in anti-phase and, as a result, the magnetization of the 
metamaterial can be locally modified \cite{Lazarides2006,Eleftheriou2008}.
Breathers exhibiting in-phase oscillations may be however generated in SQUID
metamaterials where the on-site potential may have multiple minima. 
Fundamental dissipative breathers in metallic and SQUID metamaterials are
single-sited that however cannot exist in ${\cal PT}$ metamaterials, due to the 
${\cal PT}$ symmetry. In the latter, the fundamental breather occupies at least
two sites, i.e., a dimer, which is ${\cal PT}-$symmetric by itself.  
Although experimental observations of breathers in metamaterials are still lacking,
the advances in fabrication of active and superconducting metamaterials may provide
structures with significantly reduced losses. Then, breather observation would be
in principle possible with the dynamic approaches presented above.

\begin{acknowledgement}
This work was partially supported by
the European Union's Seventh Framework Programme (FP7-REGPOT-2012-2013-1)
under grant agreement n$^o$ 316165, and
by the Thales Projects ANEMOS and MACOMSYS, co‐financed by the European Union
(European Social Fund – ESF) and Greek national funds through the Operational
Program "Education and Lifelong Learning" of the National Strategic Reference
Framework (NSRF) ‐ Research Funding Program: THALES.
Investing in knowledge society through the European Social Fund.
\end{acknowledgement}

\end{document}